\documentclass[runningheads]{llncs}
\usepackage[T1]{fontenc}
\usepackage{graphicx}
\usepackage{amsmath,amssymb,amsfonts}
\usepackage{textcomp}
\usepackage{tabularx}
\usepackage{booktabs}
\usepackage{array}
\usepackage{xcolor}
\usepackage[table,xcdraw]{xcolor}
\usepackage{colortbl}
\newcolumntype{Y}{>{\centering\arraybackslash}X}
\newcolumntype{P}[1]{>{\centering\arraybackslash}p{#1}}

\begin{document}
\title{A Comprehensive Review of Advancements in Powering and Charging Systems for Unmanned Aerial Vehicles}
\titlerunning{A Review of Powering and Charging Systems for UAVs}
%
\author{Harsh Abhinandan\inst{1}\orcidID{0000-0001-1111-2222} \and
Aditya Dhanraj\inst{1}\orcidID{1111-2222-3333-4444} \and
Aryan Katoch\inst{1}\orcidID{2222-3333-4444-5555} \and
R. Raja Singh\inst{2}\orcidID{3333-4444-5555-6666}}
\authorrunning{H. Abhinandan et al.}
%
\institute{School of Electrical and Electronics Engineering, Vellore Institute of Technology, Vellore, India\\
\email{\{harshabhi123,dhanrajaditya247,aryankatoch909\}@gmail.com} \and
Automotive Research Centre, Vellore Institute of Technology, Vellore, India\\
\email{rrajasingh@vit.ac.in}}
\maketitle              
\begin{abstract}
Unmanned Aerial Vehicles (UAVs) or drones have witnessed a spectacular surge in applications for military, commercial, and civilian purposes. However, their potential for flight is always limited by the finite power budget of their onboard power supplies. The limited flight time problem has led to intensive research into new sources of power and innovative charging strategies to enable protracted, autonomous flight. This paper gives a comparative summary of the current state-of-the-art in UAV power and refuelling technology. The paper begins with an analysis of the variety of energy sources, from classical batteries to fuel cells and hybrid systems, based on their relative advantages and disadvantages in energy density, weight, and safety. Subsequently, the review explores a spectrum of replenishment options, from simple manual battery swapping to sophisticated high-tech automatic docking stations and smart contact-based charging pads. Most of the review is dedicated to the newer technology of wireless power transfer, which involves near-field (inductive, capacitive) and far-field (laser, microwave) technology. The article also delves into the most important power electronic converter topologies, battery management systems, and control approaches that form the core of these charging systems. Finally, it recapitulates the most significant challenges in technical, economic, and social aspects for promising avenues of future research. The comprehensive review is a valuable guide for researchers, engineers, and policymakers striving to enhance UAV operational performance.

\end{abstract}
\section{Introduction}
\begin{figure}
    \centering
    \includegraphics[width=0.8\linewidth]{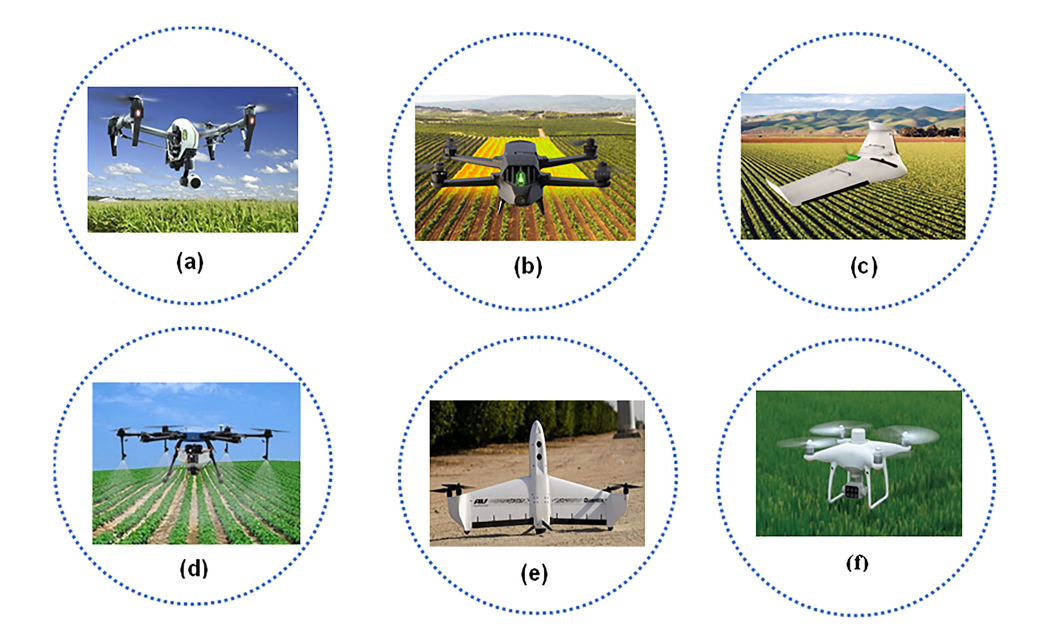}
    \caption{Illustrative representation of various UAV types deployed across multiple domains, including agriculture, logistics, and surveillance applications \cite{ref_bongomin2024uav}.}
    \label{fig:uav}
\end{figure}

Unmanned Aerial Vehicles have evolved from being specialised weapons for the military to ubiquitous assets across industries ranging from agriculture and logistics to surveillance and search and rescue as shown in Figure \ref{fig:uav}. Their agility, compact nature, and high maneuverability render them ideally suited for an incredibly wide range of applications. The primary limitation to their widespread, extended application is the low discharge and capacity characteristics of onboard power sources, typically lithium-ion batteries \cite{ref_obaidi2018,ref_Voznesenskii}. The flight durations of most commercial rotor-based UAVs are restricted to approximately 20-40 minutes and subsequently must land to be recharged or have their batteries swapped \cite{ref_fetisov}. This limitation severely restricts mission time and operating distance, with the necessity for human intervention and inducing unnecessary downtime.

Thereafter, the development of robust, autonomous, and effective charging and energy solutions has become a focal area of research, to solve the issue of flight endurance and achieve the full potential of UAV technology \cite{ref_rong2023,ref_kodeeswaran2025comparative}. This review provides a comprehensive overview of the prime sources of power, recharging methods, and emerging technologies that will allow for long-endurance UAV flight. Their low operating costs and high mobility have also increased their demand for integration into large technological ecosystems like the Internet of Things (IoT). UAVs are no longer just end devices in such scenarios but can also operate as dynamic, aerial infrastructure. These UAV-based services are categorised into four main groups: data-related services (acting as mobile collectors), charging batteries of ground sensors, relaying communications, and providing Mobile Edge Computing (MEC) capabilities \cite{ref_pakrooh2021survey}.

\begin{figure*}
    \centering
    \includegraphics[width=0.8\linewidth]{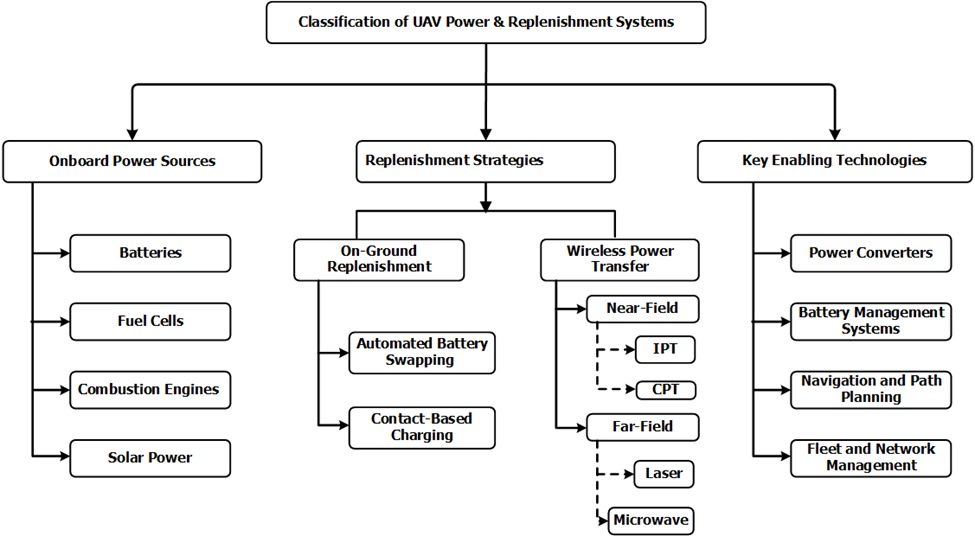}
    \caption{Perspective flow chart.}
    \label{fig:flow_chart}
\end{figure*}

\section{Power Sources for Unmanned Aerial Vehicles}
The selection of power sources affects the flight time, payload, and overall efficiency of a UAV. The most common power sources for drones are typically categorised into batteries, fuel cells, combustion engines, and solar power as shown in Figure \ref{fig:flow_chart} \cite{ref_kodeeswaran2025comparative}.

\begin{table}[]
\centering
\caption{Comparison of UAV Power Sources.}
\label{tab:uav_power_sources}
\begin{tabular}{|l|l|l|l|l|l|l|l|}
\hline
\rowcolor[HTML]{FFFFFF} 
\multicolumn{1}{|c|}{\cellcolor[HTML]{FFFFFF}\textbf{\begin{tabular}[c]{@{}c@{}}Power \\ Source\end{tabular}}} & \multicolumn{1}{c|}{\cellcolor[HTML]{FFFFFF}\textbf{\begin{tabular}[c]{@{}c@{}}Specific \\ Energy\end{tabular}}} & \multicolumn{1}{c|}{\cellcolor[HTML]{FFFFFF}\textbf{\begin{tabular}[c]{@{}c@{}}Flight \\ Time\end{tabular}}} & \multicolumn{1}{c|}{\cellcolor[HTML]{FFFFFF}\textbf{Weight}} & \multicolumn{1}{c|}{\cellcolor[HTML]{FFFFFF}\textbf{\begin{tabular}[c]{@{}c@{}}Payload \\ Capability\end{tabular}}} & \multicolumn{1}{c|}{\cellcolor[HTML]{FFFFFF}\textbf{\begin{tabular}[c]{@{}c@{}}Recharge/ \\ Discharge \\ Time\end{tabular}}} & \multicolumn{1}{c|}{\cellcolor[HTML]{FFFFFF}\textbf{Cost}} & \multicolumn{1}{c|}{\cellcolor[HTML]{FFFFFF}\textbf{Key Demerit}}                           \\ \hline
\begin{tabular}[c]{@{}l@{}}Li-Po \\ Battery\end{tabular}                                                       & High                                                                                                             & Moderate                                                                                                     & Light                                                        & Limited                                                                                                             & Moderate                                                                                                                     & Moderate                                                   & \begin{tabular}[c]{@{}l@{}}Can't \\ sustain \\ peak \\ power \\ requirements\end{tabular}   \\ \hline
\begin{tabular}[c]{@{}l@{}}Hydrogen \\ Fuel Cell\end{tabular}                                                  & High                                                                                                             & Long                                                                                                         & Moderate                                                     & High                                                                                                                & Fast                                                                                                                         & High                                                       & \begin{tabular}[c]{@{}l@{}}Bulky, with \\ uncertain \\ operating \\ costs\end{tabular}      \\ \hline
\begin{tabular}[c]{@{}l@{}}Combustion \\ Engine\end{tabular}                                                   & Moderate                                                                                                         & Long                                                                                                         & Heavy                                                        & High                                                                                                                & Fast                                                                                                                         & High                                                       & \begin{tabular}[c]{@{}l@{}}Heavy, noisy, \\ restricted \\ to fixed wing\end{tabular}        \\ \hline
\begin{tabular}[c]{@{}l@{}}Solar \\ Panels\end{tabular}                                                        & Low                                                                                                              & \begin{tabular}[c]{@{}l@{}}Very \\ Long\end{tabular}                                                         & Heavy                                                        & Limited                                                                                                             & N/A                                                                                                                          & Moderate                                                   & \begin{tabular}[c]{@{}l@{}}Panel space \\ limitations, \\ weather \\ dependent\end{tabular} \\ \hline
\end{tabular}
\end{table}
\begin{enumerate}
    \item \textbf{Batteries:} Lithium-Polymer (Li-Po) and Lithium-Ion (Li-ion) batteries are the most used power systems for commercial UAVs because they have high power density, good dependability, and low weight-to-volume. Their biggest drawback, however, is a low energy density, which translates into short flight times, usually only 20-40 minutes \cite{ref_kodeeswaran2025comparative,ref_jiao,ref_fetisov}.
    \item \textbf{Fuel Cells (FCs):} Hydrogen fuel cells have a much higher energy density than batteries, and they may even extend flight times from minutes to hours. Proton Exchange Membrane (PEM) FCs are a strong option with an energy density up to 150 times that of a Li-Po battery. Their main shortcomings are their size, the sophistication of storing hydrogen, and higher operating expenses \cite{ref_kodeeswaran2025comparative}.
    \item \textbf{Combustion Engines:} Internal combustion engines (ICEs) are a weighty alternative, particularly for large fixed-wing drones, providing long endurance and high payload carrying capacity. They are heavy, noisy, produce smoke, and are not suited well for small multi-rotor drones \cite{ref_kodeeswaran2025comparative}.
    \item \textbf{Solar power:} The method is well suited for fixed-wing UAVs with high surface area where solar panels are installed \cite{ref_fetisov}. Photovoltaic (PV) cells offer the advantage of onboard power generation, which makes it possible to have potentially unlimited endurance for ultra-long-endurance, long-duration missions. The primary disadvantages are that they are sunlight-dependent and have low efficiency of conversion, causing a big surface requirement and making them less ideal for small multirotor drones \cite{ref_kodeeswaran2025comparative,ref_jiao}. A comparison of the sources of power is presented in Table \ref{tab:uav_power_sources}.
\end{enumerate}   

\section{On-Ground Replenishment Strategies}
To address the disadvantage of a single charge, multiple ground-based replenishment strategies have been proposed. These can mainly be divided into docking stations for battery swapping and contact charging platforms.

\subsection{Docking Stations and Automated Battery Swapping}
Battery swapping through automation is a fast solution with minimal downtime for drones. With this method, the UAV touches down at a docking station, where a robotic system swaps the spent battery with a charged one. This reduces the replenishment time to only a few minutes, allowing near-continuous use. However, these systems have complex mechanical components, which increases their price and may make them impractical for general deployment. The architecture of such stations would require the need for accurate landing and positioning mechanisms to orient the UAV so the robotic gripper can swap batteries.

\subsection{Wired/Contact-Based Charging Stations}
Contact-based charging stations present a less expensive and less complex method of battery swapping in relation to battery swapping \cite{ref_Voznesenskii}. Contact-based systems require the UAV to land on a platform where it makes physical contact with ground-based charging pads through electrical contacts, which would initiate the charging process. A primary challenge is ensuring stable electrical contact, especially with consideration of potential landing inaccuracies due to wind or GPS failure. To achieve this, strong docking stations are being engineered with both mechanical and electronic systems that provide a controlled environment. Nieuwoudt et al., for example, developed and constructed a full, self-contained automated security UAV docking station. Their setup includes a weather-protecting sliding roof and, most importantly, a 4-axis mechanical centring system. This centring system employs actuators that physically move the position of the drone once it has been placed on the platform so that charging contacts on a centring arm make a consistent contact with the UAV charging strips. In order to navigate the drone through the final, crucial landing phase, these stations typically use high-precision, close-range aids to navigation such as vision-based systems with fiducial markers (e.g., ArUco markers) or infrared (IR) beacons \cite{ref_nieuwoudt2023}.

Correspondingly, Al-Obaidi et al. proposed an entirely automatic charging station consisting of a ground platform composed of alternating positive and negative polarity square-shaped copper plates, like a chessboard. It considers changing the orientation of the UAV upon landing. The onboard circuit utilises six bridge diode rectifiers to reverse the polarity from the four contact terminals of the UAV to facilitate independent recharging regardless of any yaw angle and simplify landing operations. A second approach utilises a flat parallel electrode platform, as reported by Fetisov V and Akhmerov S. A compilation of on-ground replenishment techniques is shown in Table \ref{tab:replenishment_strategies}.

\begin{table}[]
\centering
\caption{Comparison of On-Ground UAV Replenishment Strategies.}
\label{tab:replenishment_strategies}
\begin{tabular}{|l|l|l|l|l|l|}
\hline
\rowcolor[HTML]{EFEFEF} 
\multicolumn{1}{|c|}{\cellcolor[HTML]{EFEFEF}\textbf{Strategy}}      & \multicolumn{1}{c|}{\cellcolor[HTML]{EFEFEF}\textbf{\begin{tabular}[c]{@{}c@{}}Replenishment \\ Speed\end{tabular}}} & \multicolumn{1}{c|}{\cellcolor[HTML]{EFEFEF}\textbf{\begin{tabular}[c]{@{}c@{}}System \\ Complexity\end{tabular}}} & \multicolumn{1}{c|}{\cellcolor[HTML]{EFEFEF}\textbf{\begin{tabular}[c]{@{}c@{}}Level of \\ Autonomy\end{tabular}}} & \multicolumn{1}{c|}{\cellcolor[HTML]{EFEFEF}\textbf{Cost}}  & \multicolumn{1}{c|}{\cellcolor[HTML]{EFEFEF}\textbf{Key Features}}                                                                  \\ \hline
\begin{tabular}[c]{@{}l@{}}Manual \\ Battery \\ Swap\end{tabular}    & \begin{tabular}[c]{@{}l@{}}Very Fast \\ (seconds)\end{tabular}                                                       & Low                                                                                                                & \begin{tabular}[c]{@{}l@{}}Low \\ (human \\ required)\end{tabular}                                                 & Low                                                         & \begin{tabular}[c]{@{}l@{}}Simple but \\ time-consuming.\end{tabular}                                                               \\ \hline
\begin{tabular}[c]{@{}l@{}}Automated \\ Battery \\ Swap\end{tabular} & \begin{tabular}[c]{@{}l@{}}Quick \\ (minutes)\end{tabular}                                                           & \begin{tabular}[c]{@{}l@{}}High \\ (Mechanical)\end{tabular}                                                       & High                                                                                                               & High                                                        & \begin{tabular}[c]{@{}l@{}}Continuous \\ operation\\ complicated and \\ expensive.\end{tabular}                                     \\ \hline
\begin{tabular}[c]{@{}l@{}}Contact-\\ Based \\ Charging\end{tabular} & \begin{tabular}[c]{@{}l@{}}Sluggish \\ (hours)\end{tabular}                                                          & \begin{tabular}[c]{@{}l@{}}Moderate \\ (Electrical)\end{tabular}                                                   & High                                                                                                               & Moderate                                                    & \begin{tabular}[c]{@{}l@{}}Less complicated \\ mechanics; \\ needs accurate \\ landing or \\ strong contact \\ design.\end{tabular} \\ \hline
\begin{tabular}[c]{@{}l@{}}Wireless \\ Charging\end{tabular}         & \begin{tabular}[c]{@{}l@{}}Moderate \\ to Slow\end{tabular}                                                          & \begin{tabular}[c]{@{}l@{}}Moderate \\ (Electronics)\end{tabular}                                                  & \begin{tabular}[c]{@{}l@{}}Very \\ High\end{tabular}                                                               & \begin{tabular}[c]{@{}l@{}}Moderate \\ to High\end{tabular} & \begin{tabular}[c]{@{}l@{}}High convenience, \\ no physical contact; \\ sensitive \\ to misalignment.\end{tabular}                  \\ \hline
\end{tabular}
\end{table}

\begin{figure}
    \centering
    \includegraphics[width=0.9\columnwidth]{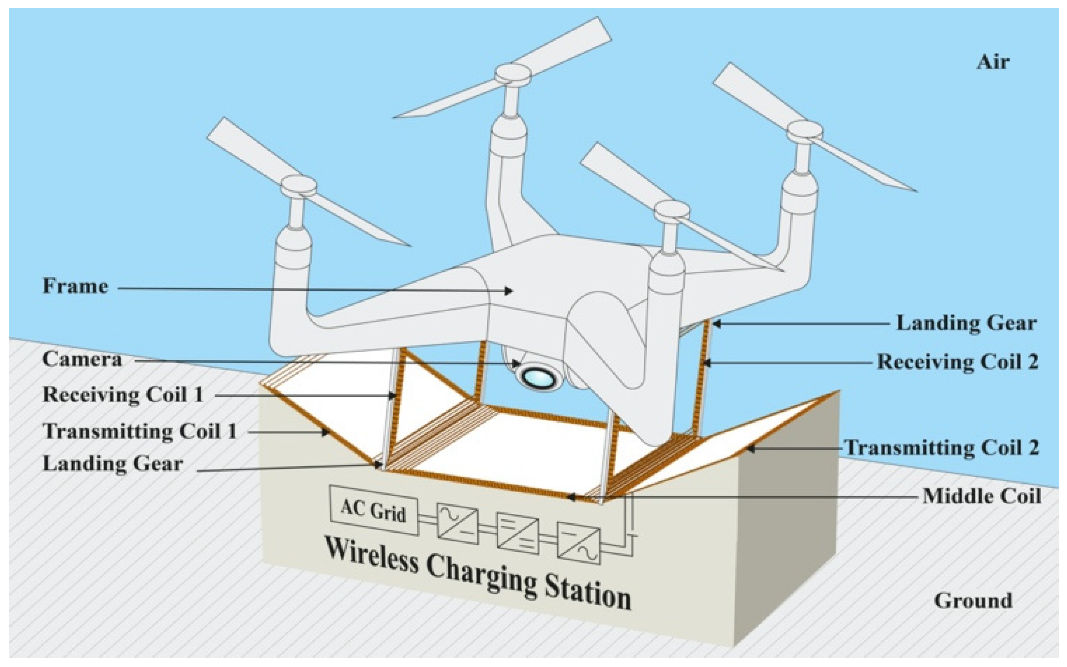}
    \caption{An illustrative diagram of wireless power transfer for drones, showing a quadcopter equipped with receiving coils aligned over a wireless charging station with transmitting coils for inductive energy transfer \cite{ref_augccal2024novel}.}
    \label{fig:ipt_charging_station}
\end{figure}

\section{Wireless Power Transfer for UAVs}
Wireless Power Transfer (WPT) technology has been considered as a promising solution, providing convenience in charging without physical connectors, thus increasing automation and flexibility as shown in Figure \ref{fig:ipt_charging_station}. WPT can be divided into near-field and far-field technologies, with a comparative overview of each presented in Table \ref{tab:wpt_technologies}.
\begin{table}[]
\centering
\caption{Overview of Wireless Power Transfer (WPT) Technologies for UAVs.}
\label{tab:wpt_technologies}
\begin{tabular}{|l|l|l|l|l|l|}
\hline
\rowcolor[HTML]{EFEFEF} 
\multicolumn{1}{|c|}{\cellcolor[HTML]{EFEFEF}\textbf{\begin{tabular}[c]{@{}c@{}}WPT \\ Technology\end{tabular}}} & \multicolumn{1}{c|}{\cellcolor[HTML]{EFEFEF}\textbf{Principle}}       & \multicolumn{1}{c|}{\cellcolor[HTML]{EFEFEF}\textbf{Range}}      & \multicolumn{1}{c|}{\cellcolor[HTML]{EFEFEF}\textbf{Efficiency}} & \multicolumn{1}{c|}{\cellcolor[HTML]{EFEFEF}\textbf{\begin{tabular}[c]{@{}c@{}}Power \\ Level\end{tabular}}} & \multicolumn{1}{c|}{\cellcolor[HTML]{EFEFEF}\textbf{Key Challenge(s)}}                       \\ \hline
\begin{tabular}[c]{@{}l@{}}Inductive \\ (IPT)\end{tabular}                                                       & \begin{tabular}[c]{@{}l@{}}Magnetic \\ Field \\ Coupling\end{tabular} & \begin{tabular}[c]{@{}l@{}}Near-field \\ (cm to m)\end{tabular}  & \begin{tabular}[c]{@{}l@{}}High \\ ($>$90\%)\end{tabular}        & \begin{tabular}[c]{@{}l@{}}High \\ (kW)\end{tabular}                                                         & \begin{tabular}[c]{@{}l@{}}Misalignment \\ sensitivity, \\ weight\end{tabular}               \\ \hline
\begin{tabular}[c]{@{}l@{}}Capacitive \\ (CPT)\end{tabular}                                                      & \begin{tabular}[c]{@{}l@{}}Electric \\ Field \\ Coupling\end{tabular} & \begin{tabular}[c]{@{}l@{}}Near-field \\ (mm to cm)\end{tabular} & \begin{tabular}[c]{@{}l@{}}Moderate-\\ High\end{tabular}         & \begin{tabular}[c]{@{}l@{}}Moderate-\\ High\end{tabular}                                                     & \begin{tabular}[c]{@{}l@{}}Small air gap, \\ misalignment \\ sensitivity\end{tabular}        \\ \hline
\begin{tabular}[c]{@{}l@{}}Laser \\ (LPT)\end{tabular}                                                           & \begin{tabular}[c]{@{}l@{}}Light \\ Beam\end{tabular}                 & \begin{tabular}[c]{@{}l@{}}Far-field \\ (m to km)\end{tabular}   & \begin{tabular}[c]{@{}l@{}}Low \\ ($<$15\%)\end{tabular}         & \begin{tabular}[c]{@{}l@{}}High \\ (kW)\end{tabular}                                                         & \begin{tabular}[c]{@{}l@{}}Safety concerns, \\ atmospheric \\ attenuation, cost\end{tabular} \\ \hline
\begin{tabular}[c]{@{}l@{}}Microwave \\ (MPT)\end{tabular}                                                       & \begin{tabular}[c]{@{}l@{}}Radio \\ Waves\end{tabular}                & \begin{tabular}[c]{@{}l@{}}Far-field \\ (m to km)\end{tabular}   & \begin{tabular}[c]{@{}l@{}}Very Low \\ ($<$10\%)\end{tabular}    & \begin{tabular}[c]{@{}l@{}}Moderate-\\ High\end{tabular}                                                     & \begin{tabular}[c]{@{}l@{}}Low efficiency, \\ safety, interference\end{tabular}              \\ \hline
\end{tabular}
\end{table}
\subsection{Near-Field WPT}
Near-field WPT is used over short ranges and is characterised by Inductive Power Transfer and Capacitive Power Transfer.

\begin{enumerate}
    \item \textbf{Inductive Power Transfer (IPT):} IPT utilises magnetic fields generated by coils to conduct energy from a ground transmitter to an UAV receiver \cite{ref_kodeeswaran2025comparative,ref_kamble}. IPT can be highly efficient but is extremely vulnerable to alignment between receiver and transmitter coils, an issue that is still a major challenge for dynamic UAV landings \cite{ref_danciu}. A systematic review by Mou et al. identifies that the primary engineering challenge is the compromise between performance and the weight gain of the onboard system. The essence of the paper is in its classification of technical solutions, with current studies being geared towards the design of sophisticated coupler configurations (e.g., planar coil arrays, frustum-shaped chargers, and helical coils) and advanced compensation topologies (e.g., LCL and LCC circuits) to generate more homogeneous magnetic fields and enhance tolerance to both lateral and angular misalignment. Its performance decreases significantly as the distance grows and as there is misalignment between the coils, a significant disadvantage for UAV uses where accurate landing cannot always be assured \cite{ref_Voznesenskii,ref_rong2023,ref_mou2023}.
    \item \textbf{Capacitive Power Transfer (CPT):} CPT takes advantage of the electric field between conductive plates to transfer power. It has advantages such as lower eddy current losses and potentially simpler, lighter construction than IPT. Its primary advantage is its potential for lower-weight, lower-cost couplers with reduced susceptibility to Eddy current losses and EMI. Yet, its main drawback is that it usually needs a significantly smaller air gap (on the order of millimeters) and is extremely sensitive to variations in said gap, which makes it not easy for applications where landing height varies \cite{ref_mou2023}.
\end{enumerate}

\subsection{Far-Field WPT and New Ideas}
\begin{enumerate}
    \item \textbf{Laser Power Transfer (LPT):} It employs a high-energy laser beam to transfer power from a ground station to a photovoltaic receiver onboard the UAV. It is feasible for long-range, in-flight recharging but is costly and involves great safety hazards \cite{ref_obaidi2018,ref_fetisov}.
    \item \textbf{Microwave Power Transfer (MPT):} MPT utilizes microwaves to transfer power from a ground antenna to a rectenna mounted on the UAV. The technology is omnidirectional but is prone to environmental attenuation and has low efficiency \cite{ref_obaidi2018,ref_rong2023}.
    \item \textbf{RF Energy Transfer and SWIPT:} In the application of UAV-enabled IoT, the function of the UAV may be switched from energy consumption to energy supply. As surveyed by Pakrooh \& Bohlooli, UAVs may employ Radio Frequency (RF) signals to wirelessly recharge low-power ground sensor batteries. The idea of this concept lies in prolonging the operational lifetime of a whole sensor network. This is then further developed into Simultaneous Wireless Information and Power Transfer (SWIPT), wherein a single RF signal is smartly divided at the sensor node to provide power to its circuitry while simultaneously transferring its data payload \cite{ref_pakrooh2021survey}.
    \item \textbf{Drone-to-Drone Charging:} One concept of the future is moving the charging infrastructure away from a fixed ground station to a dynamic, mobile station. The crux of the Anand et al. proposed system is an autonomous two-drone design: a main "mission drone" and a support "charging drone" with a wireless power transmitter. The charging drone would find, dock with, and recharge the main drone on its own, either in flight or at a temporary landing area. This tactic could greatly improve mission duration for specialised missions like Electronic Warfare Support, where returning to a fixed location is impossible \cite{ref_tp2025design}.
\end{enumerate}

\section{Critical Technologies and Topologies in Charging Systems}
The efficiency of any charging system, either wired or wireless, critically depends on the power electronics underlying it, i.e., the converter topologies and their control systems.

\subsection{Converter Topologies}
Efficient power conversion is essential in minimising charging time and energy loss. Synchronous buck converters are used preferentially over conventional buck converters because they replace the diode with a MOSFET, having a very low on-state resistance, thereby enhancing efficiency. Interleaved synchronous buck converters are utilised in high-power applications \cite{ref_singh2024,ref_swathi}. For the integration of battery storage in high-voltage systems, multilevel converter topologies like the Modular Multilevel Converter (MMC) are ideal \cite{ref_krishna2025}.

\subsection{Battery Management Systems (BMS)}
The functionality, safety, and life of a UAV power system highly rely on the Battery Management System (BMS). A review by Jiao et al. considers the BMS a key element with duties beyond basic charging. These operations are classified into three broad categories:
\begin{enumerate}
    \item Discharging and charging control, which involves executing charging strategies (for example, CC-CV) and battery equalisation to ensure all cells of a pack are equalised.
    \item Estimation of the battery state, which means precisely estimating the State of Charge (SOC), State of Health (SOH), and Remaining Useful Life (RUL).
    \item System safety, which entails data protection, fault diagnosis, and thermal management. A smart BMS is central to maximising battery performance as well as the overall safety of the UAV \cite{ref_jiao}.
\end{enumerate}

\subsection{Path Planning and Navigation to Charging}
Autonomous navigation to a charging point by a drone would necessitate advanced path planning and navigation algorithms. In GPS-enabled environments, this is easy. For operations in GPS-denied or complex environments, advanced techniques are necessary. Systems utilise Simultaneous Localisation and Mapping (SLAM) to build a real-time map of the environment with sensors such as LiDAR or cameras \cite{ref_tp2025design}. Having a map and/or knowledge of the destination, pathfinding algorithms such as Dijkstra's or A* are employed to calculate the shortest collision-free path to the charging station, allowing dynamic re-routing to circumvent unexpected obstacles \cite{ref_tp2025design}.

\subsection{Fleet and Network Management}
In large-scale operations of UAVs, for example, in telemedicine delivery, fleet management of drones and network management of charging stations is a highly complex logistics issue. Danciu et al. suggest a "charging network" that is tasked with routing the UAVs to the closest available station, scheduling the charging time, and collecting statistics for optimising the system. This network-oriented strategy is necessary to prevent congestion at charging stations and to make sure that mission-critical UAVs (for example, ones transporting medical supplies) are given priority, thus ensuring optimum efficiency of the entire fleet \cite{ref_danciu}.

\section{Challenges, Gaps in Research, and Future Directions}
The journey toward completely autonomous UAV flight is characterized by noteworthy challenges and promising lines of research. One key area includes making WPT more robust and efficient. Existing near-field WPT solutions are severely affected by landing misalignment, whereas far-field technology, although it has the advantage of charging in-flight, has low efficiency and is a safety risk. Future developments would need to involve creating new coil designs, sophisticated compensation topologies, and smart beam-steering/tracking systems, along with rigorous safety standards and regulatory cooperation.

A second significant barrier is reducing the Size, Weight, Power, and Cost (SWaP-C) burden of onboard charging systems. This means moving toward miniaturization, clean airframe integration, and employing composites and wide-bandgap semiconductors (such as Gallium Nitride) to develop power electronics that are smaller and more efficient. Developing multifunctional structures where charging components are integrated right into the airframe will help minimize SWaP-C.

Beyond technical aspects, public acceptance, safety, and regulatory frameworks are crucial. Concerns about privacy, safety, and potential misuse of drones highlight a socio-technical gap. Future research should adopt an interdisciplinary approach, combining engineering with social sciences to develop "human-friendly" systems that address public concerns and proactively engage with regulators to establish robust safety and data-handling policies.

In the future, UAV power systems have a bright future ahead with intelligent integration of these technologies. This involves active docking stations with self-alignment capabilities, smarter power converters, and AI/machine learning-based Battery Management Systems (BMS). These smart BMS will manage charging cycles for optimum efficiency, forecast battery condition, and provide advanced fleet management, routing drones to optimum charging locations based on mission requirements. Ultimately, the vision is to develop a single, intelligent ecosystem where charging is an active and strategic part of the mission, enabling longer autonomous flight.

\section{Conclusion}
This paper has provided a comprehensive review of the status of UAV power and recharge systems. It illustrates that, rather than a single game-changer technology, it is a multi-faceted, application-driven solution that is pushing towards extended autonomy. The choice of any power and resupply strategy is a complex compromise between flight duration, cost, operational independence, and safety. Although lithium batteries remain the standard, the review notes that alternative power sources like hydrogen fuel cells are a revolutionary change in longevity. Furthermore, the replenishment strategies have evolved significantly, with Wireless Power Transfer (WPT) being the most technologically competent to enable an age of unattended and uninterrupted flight by drones. In addition, the review has highlighted the importance of critical enabling technologies, such as the sophisticated power electronics and smart Battery Management Systems (BMS) that are critical to the safety, efficiency, and longevity of any charging system.

This review has uncovered several of the most important research gaps, including enhancing WPT robustness to misalignment, ensuring the safety and practicality of in-flight charging, minimizing the SWaP-C burden of on-board systems, and developing scalable fleet management solutions. Most critically, there is a socio-technical gap between technology capability and public acceptance. Our suggestions for future research are thus explicitly targeted at those very areas. We propose the imperative of building robust, misalignment-resistant WPT systems; driving ahead intelligent, network-aware management systems that view fleets of UAVs and charging stations as an integrated, optimised infrastructure; and looking for multi-disciplinary research to develop systems that are technology leadership, safety, regulation, and society-trusting. In the end, closing the gap between laboratory-validated concepts and strong, real-world deployment will be the sign of success. Addressing these problems will be necessary to unleash the full potential of autonomous UAV action during the next several decades.

%
%
%
%

\end{document}